\documentclass[onecolumn,amsmath,amssymb]{revtex4}
\usepackage{bm}
\usepackage{graphicx}
\usepackage{epstopdf}
\usepackage{amsmath}
\usepackage{lscape}

\usepackage[T1]{fontenc}
\usepackage[polish,english]{babel}
\usepackage[utf8]{inputenc}

\begin{document}

\newcommand{\uu}[1]{\underline{#1}}
\newcommand{\pp}[1]{\phantom{#1}}
\newcommand{\be}{\begin{eqnarray}}
\newcommand{\ee}{\end{eqnarray}}
\newcommand{\ve}{\varepsilon}
\newcommand{\vs}{\varsigma}
\newcommand{\Tr}{{\,\rm Tr\,}}
\newcommand{\pol}{\frac{1}{2}}
\newcommand{\RR}{\rotatebox[origin=c]{180}{$\mathbb{R}$} }
\newcommand{\CC}{\rotatebox[origin=c]{180}{$\mathbb{C}$} }
\newcommand{\rr}{\mathbb{R}}
\newcommand{\Exp}{{\,\rm Exp\,}}
\newcommand{\Sin}{{\,\rm Sin\,}}
\newcommand{\Cos}{{\,\rm Cos\,}}
\newcommand{\Sinh}{{\,\rm Sinh\,}}
\newcommand{\Cosh}{{\,\rm Cosh\,}}

\title{
Swapping space for time: An alternative to time-domain interferometry}
\author{Marek Czachor}
\affiliation{
Katedra Fizyki Teoretycznej i Informatyki Kwantowej,
Politechnika Gda\'nska, 80-233 Gda\'nsk, Poland
}

\begin{abstract}
Young's double-slit experiment \cite{Young} requires two waves produced simultaneously at two different points in space.  In quantum mechanics the waves correspond to a single quantum object, even as complex as a big molecule. An interference is present as long as one cannot tell for sure which slit is chosen by the object. The more we know about the path, the worse the interference. 
In the paper we show that quantum mechanics allows for a dual version of the phenomenon: self-interference of waves propagating through a single slit but at different moments of time. The effect occurs for time-independent Hamiltonians and thus should not be confused with Moshinsky-type time-domain interference \cite{Moshinsky}, a consequence of active modulation of parameters of the system (oscillating mirrors, chopped beams, time-dependent apertures, moving gratings, etc.).  The discussed phenomenon is counterintuitive even for those who are trained in quantum interferometry. For example, the more we know about the trajectory in space, the better the interference. Exactly solvable models lead to formulas deceptively similar to those from a Youngian analysis.  There are reasons to believe that this new type of quantum interference  was already observed in atomic interferometry almost three decades ago, but was misinterpreted and thus rejected as an artifact of unknown origin. 
\end{abstract}
\maketitle

\section{Introduction}

The idea of time-domain interferometry can be traced back to the seminal paper by Moshinsky \cite{Moshinsky} on diffraction in time. Since then, the issue was both theoretically and experimentally investigated by a number of authors. With apparently no exception, all the examples discussed in the literature were based on time-dependent Hamiltonians (a time-dependent magnetic field\cite{Badurek86}, a periodically opened grating \cite{Hill}, a moving grating \cite{Balashov}, a moving mirror \cite{Arndt,Szriftgiser}, laser-controlled windows of short duration \cite{Lindner}). Different combinations of interference in space and time were also discussed \cite{Bruckner}. A review of the first five decades of research in the field can be found in \cite{PB}. 

In what follows, I discuss a new quantum interferometric phenomenon, in some respects similar to interference in time. Its manifestations can be confused with Young's double-slit interference. They may be at the heart of a controversy raised by an experiment performed almost three decades ago.

\begin{figure}
\includegraphics[width=8 cm]{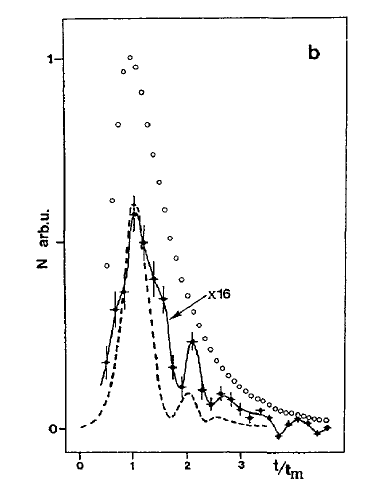}
\caption{An example of the data from \cite{1992}. Intensity of the emitted radiation as a function of time of flight $t$ (in units of the most probable time of flight $t_m$). The dashed line is an analogous prediction for an appropriate Young-type single-photon experiment.}
\label{exp}
\end{figure}

In 1991-1992 the atom interferometer group from Univerist\'e Paris-Nord reported observation of an optical phenomenon that resembled a Young-type self-interference of photons spontaneously emitted from a multi-peaked atomic center-of-mass wave packet \cite{1991,1992}. The result suggested that a single-atom wave packet can play a role of a multi-point coherent source of light.  The effect was weak but clearly visible, with small error bars, and reappeared in various configurations of the experiment (Fig.~\ref{exp}). 
However, there was a fundamental problem with the data. In principle, after (or before) having detected a photon one could perform a direct measurement of the atomic position, revealing location of the source at the moment of emission, and thus destroying the interference. Accordingly, this was not a typical Young experiment, but rather its which-way version \cite{ww,Chapman}.
What one expects is a radiation typical of several incoherent sources of light. An atomic wave packet of the form $\alpha|1\rangle+\beta e^{i\phi}|2\rangle$ should emit light whose intensity contains contributions proportional to $|\alpha|^2$ and $|\beta|^2$, but Youngian terms involving $\cos\phi$ should be absent. Quantum optical analysis of spontaneous emission from extended wave packets did not support the data either  \cite{RZ,S1,S3,S4,S6,S7}. The extensive review \cite{rev} on optics and atomic interferometry did not even mention the effect. 

Yet, the plot from Fig.~\ref{exp} is very disturbing for a theorist. Is it possible that we overlook something? Is the analogy to which-way measurements as superficial as the one to Young's double slits?  And indeed, we will see that the analogy to which-way experiments may be misleading. Examples will be given of exactly solvable quantum mechanical models inheriting all the basic physical properties of the Paris experiment, but leading to predictions that may be easily confused with Young's interference. The discussed phenomenon is counterintuitive even for those who are trained in entangled-state interferometry. One of its possible interpretations is in terms of an interference of fields emitted from the same point in space but at different moments of time. The emitted radiation may contain contributions proportional to  $\cos\phi$, even though the initial atomic state is $\alpha|1\rangle+\beta e^{i\phi}|2\rangle$. 

\begin{figure}
\includegraphics[width=8 cm]{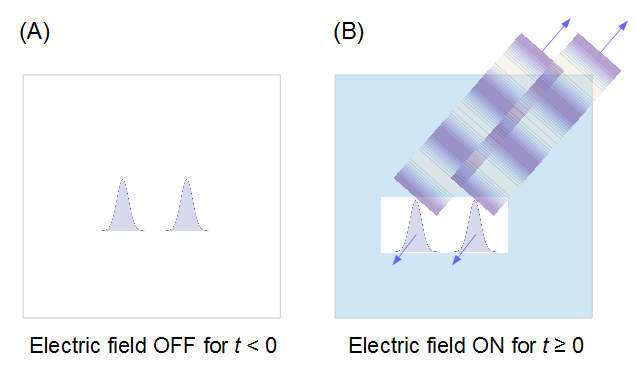}
\caption{Standard configuration analyzed in the literature. At $t=0$ all the parts of the wave packet start to decay. The act of emission changes atomic center-of-mass momentum via a recoil, creating an entangled atom-photon state. Tracing out the atomic degrees of freedom one obtains a mixed state of a single photon, with incoherent contributions from both atomic peaks. Spectrum of the emitted radiation is modified only by Doppler shifts, statistically distributed in accordance with the probability density of atomic center-of-mass  momenta.}
\label{Fig-1}
\end{figure}

\begin{figure}
\includegraphics[width=8 cm]{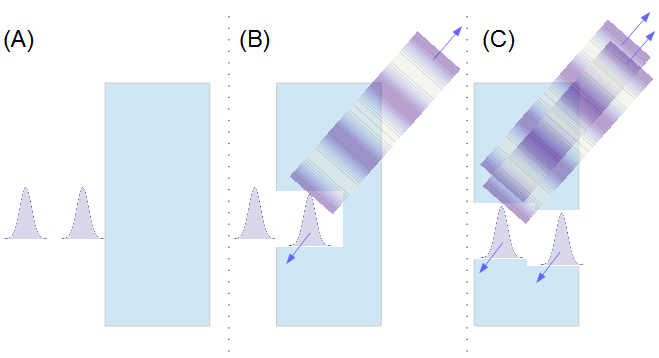}
\caption{Configuration closer to the Paris experiment. Different parts of the wave packet couple to the external field in a position dependent way. In principle, one can observe interference of photons emitted from the same point in space, but at different moments of time. Here standard intuitions from entangled-state interferometry fail, and an exact quantum mechanical solution is needed.}
\label{Fig-2}
\end{figure}

In order to understand the problem let us have a look at Fig.~\ref{Fig-1} describing the conceptual  structure of standard theoretical papers (an exception is \cite {CY}). The analyzed radiation pattern is obtained under the assumption that at $t=0$ the spontaneous emission is `turned on' in a position independent way, but the experiment was closer to the scheme from Fig.~\ref{Fig-2}. A beam of hydrogen atoms was prepared in a metastable 2s$_{1/2}$ internal state in such a way that its center-of-mass wave function consisted of three well separated (by 120 or 240 nm) peaks. The atoms moved with average velocity 10 km/s toward a region of 40 V/cm static electric field. Stark mixing $2s_{1/2}\leftrightarrow 2p_{1/2}$ followed by spontaneous emission $2p_{1/2}\to 1s_{1/2}$ of Lyman-$\alpha$ 121.6 nm photons was induced in the interaction zone. The observed dependence of intensity of radiation on the shape of the atomic wave function was compared with an analogous prediction for the Young experiment, where instead of a multi-peaked source one took a multi-hole obstacle. The resulting intensity curves were qualitatively similar.

The authors of \cite{1992} were well aware of the theoretical difficulties one will encounter in a realistic modeling of their experiment. They wrote: ``In conclusion it seems that an interference phenomenon characterized by a wavelength close to the Lyman $\alpha$ one does occur in the optical emission. The theoretical interpretation of this experiment is a priori rather difficult, not only because of the emission process itself but also because of the complexity of the present induced emission process (2s-2p transition in a profile of electric field combined with the 2p-1s transition).'' 

An early attempt of including non-simultaneity of excitation in a Weisskopf-Wigner approach can be found in \cite{CY}, but the results were difficult to interpret due to a large number of uncontrollable approximations. The problem is so fundamental that it would be imprudent to base conclusions on  approximate results. Let us note, however, that the essence of quantum self-interference can be discussed already in a two-dimensional Hilbert space. The main counterargument against the very possibility of self-interference in a which-way experiment can be formulated with only two qubits, hence in four dimensions. In spite of low dimensionality, the available formal structures are there rich enough for proof-of-principle conclusive arguments. So, it is best to follow an analogous strategy. Striping the problem of unnecessary details we have to maintain certain physical characteristics:
\begin{enumerate}
\item The position state outside of the interaction zone should be a superposition of at least two orthogonal states, $|1\rangle$ and $|2\rangle$.
\item While the system is outside of the interaction zone the field should be in a vacuum state $|0\rangle$, and the internal state of the system should be excited $|+\rangle$.
\item One should be able to distinguish between states inside and outside of the interaction zone, so there must exist at least one position state $|3\rangle$, corresponding to the region of space where spontaneous emission occurs. This state has to be orthogonal to both $|1\rangle$ and $|2\rangle$.
\item The system should be able to propagate into the interaction zone, so its free Hamiltonian must be nontrivial.
\item The emitted state of light should be orthogonal to the vacuum, but a single radiated state $|\bm k\rangle=|1\rangle$ is enough. The emitted particle can be bosonic, fermionic, or whatever, since identical formal problems occur if one replaces spontaneous emission by ionization, or by any other kind of unitarity-preserving interaction.
\item The act of spontaneous emission should change the atomic state by recoil. The realistic case 
$$|0\rangle\otimes |+,\bm P\rangle \to |\bm k\rangle\otimes |-,\bm P-\bm k\rangle $$
can be replaced by
$$|0\rangle\otimes |+,3\rangle \to |1\rangle\otimes |-,3\rangle, $$
since the pair $ |+,3\rangle$, $ |-,3\rangle$, is as orthogonal as $ |+,\bm P\rangle$, $|-,\bm P-\bm k\rangle$, and it is the orthogonality of the two states that counts in the formal argument. Their exact mathematical representation is irrelevant.
\item Total Hamiltonian should be time independent, to avoid confusion with time-domain interferometry, based on time dependent Hamiltonians.
\item The emitted states should involve exclusively single particles, to avoid confusion with intensity interferometry, based on Hanbury-Brown--Twiss effect. 
\item The dynamics of the whole atom-field system must be unitary and exactly solvable.
\end{enumerate}
A Hilbert space that satisfies all these postulates is at least 12-dimensional.  The question is: Can the probability of finding  $|\bm k\rangle=|1\rangle$ depend on $\cos\phi$, if the initial state of the whole system is
$|0\rangle\otimes\big(\alpha|+,1\rangle +\beta e^{i\phi}|+,2\rangle\big)$? The answer is in the affirmative. The effect may look like a Youngian interference but its physical meaning is different. The result is generic and should be observable in a large variety of quantum systems.

\begin{figure}
\includegraphics[width=8 cm]{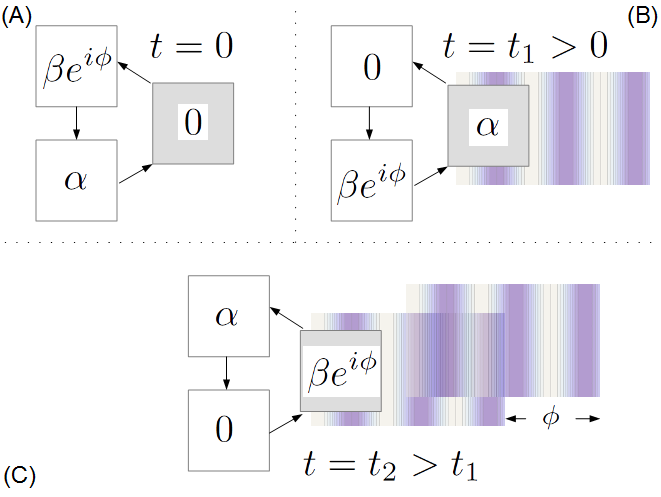}
\caption{An intuitive picture of a three-state analogue of a longitudinal Stern-Gerlach interferometer. The interaction is controlled by position $X_3$ (the gray square). The free evolution performs an anticlockwise rotation. (A) At $t=0$ the system is prepared in a two-peaked superposition of center-of-mass positions $X_1$ and $X_2$, located outside of the interaction zone. (B) At $t=t_1$ a half of the wave packet  is already in the interaction zone and emits a photon. (C) At $t=t_2$ the first half of the wave packet has already left the interaction region, but now the second half interacts with the field. In spite of its idealization, the picture correctly describes the structure of an exact quantum mechanical prediction.}
\label{Fig0}
\end{figure}

\section{Finite-state analogue of a longitudinal Stern-Gerlach interferometer}

Consider a system (`an atom') whose center of mass can occupy one of the three positions, $X_1$, $X_2$, or $X_3$, corresponding to the following three eigenstates of the discrete center-of-mass position operator $\hat X=\sum_{j=1}^3 X_j|X_j\rangle\langle X_j|$, as shown in Fig.~\ref{Fig0},
\be
|X_1\rangle
&=&
\left(
\begin{array}{c}
1\\
0\\
0
\end{array}
\right),
\quad
|X_2\rangle
=
\left(
\begin{array}{c}
0\\
1\\
0
\end{array}
\right),
\quad
|X_3\rangle
=
\left(
\begin{array}{c}
0\\
0\\
1
\end{array}
\right).
\ee
Let the two `photon' states (a vacuum or a single particle) be represented by a qubit,
\be
|0\rangle
&=&
\left(
\begin{array}{c}
1\\
0\end{array}
\right),
\quad
|1\rangle
=
\left(
\begin{array}{c}
0\\
1\end{array}
\right).
\ee
The atom is two-level,
\be
|-\rangle
&=&
\left(
\begin{array}{c}
1\\
0\end{array}
\right),
\quad
|+\rangle
=
\left(
\begin{array}{c}
0\\
1\end{array}
\right).
\ee
For simplicity we assume that in the absence of interactions the two internal atomic states have the same energy, so that we can ignore their contribution to the free Hamiltonian (for a justification of this assumption in the context of atomic interferometry see \cite{Knight}).
As required, the Hilbert space is $2\times 2 \times 3=12$ dimensional, with the basis
\be
|n,\pm,j\rangle=|n\rangle\otimes |\pm\rangle\otimes |X_j\rangle, \quad n=0,1,\, j=1,2,3
\ee
Let an `electric field' $E(\hat X)$ evaluated at the center-of-mass position satisfy $E(X_1)=E(X_2)=0$, $E(X_3)=\omega_1$, so $E(\hat X)=
\sum_{j=1}^3 E(X_j)|X_j\rangle\langle X_j|=\omega_1 |X_3\rangle\langle X_3|$. An internal `dipole moment' is $\hat d=|+\rangle\langle-|+|-\rangle\langle+|=\sigma_x$. The interaction term is taken in the usual form 
\be
\Omega_1 &=& \sigma_x\otimes \hat d\otimes E(\hat X)=\omega_1\sigma_x\otimes \sigma_x\otimes |X_3\rangle\langle X_3|,
\ee
where the leftmost $\sigma_x$ is the operator that creates or annihilates photons, and we do not assume a rotating wave approximation. 
One can also write the interaction term as 
\be
\Omega_1 &=& \hat d\otimes \hat E(\hat X)=\omega_1\sigma_x\otimes \sigma_x\otimes |X_3\rangle\langle X_3|,
\ee
where the leftmost $\sigma_x$ is treated as the dipole moment, and 
\be
\hat E(\hat X)= \sigma_x\otimes E(\hat X)
\ee
is the field operator $\hat E(X)= \sigma_x\otimes E(X)$ evaluated at the center-of-mass position operator $\hat X$. All operators are independent of time since we work in the Schr\"odinger picture and the system is closed. An act of emission or absorption of a photon is always accompanied by a change $|\pm\rangle\to |\mp\rangle$ of an internal atomic state. An entanglement we will obtain in such a toy model is an analogue of the entanglement in momentum space.

In quantum information terminology the interaction term is a two-qubit NOT gate controlled by the center-of-mass position. The free Hamiltonian is chosen in a form of a generator of rotations in  three dimensions, with the rotation axis parallel to $(1,1,1)$,
\be
\Omega_0
&=&
i \omega_0 \mathbb{I}_2\otimes \mathbb{I}_2\otimes \Big(|X_1\rangle\langle X_2|+|X_2\rangle\langle X_3|+|X_3\rangle\langle X_1|\Big)
\nonumber\\
&\pp=&+\textrm{H.c.}
\ee
The free evolution operator is thus a rotation in position space around $(1,1,1)$ with frequency $\sqrt{3} \omega _0$,
\begin{widetext}
\be
U_0(t) &=& e^{-i\Omega_0t}
=
-\frac{1}{3}\mathbb{I}_2\otimes \mathbb{I}_2\otimes 
\nonumber\\
&\pp=&
\left(
\begin{array}{lll}
 -2 \cos 2\pi t/T-1 & \cos 2\pi t/T-\sqrt{3} \sin 2\pi t/T-1 & \cos 2\pi t/T+\sqrt{3} \sin 2\pi t/T-1 \\
 \cos 2\pi t/T+\sqrt{3} \sin 2\pi t/T-1 & -2 \cos 2\pi t/T-1 & \cos 2\pi t/T-\sqrt{3} \sin 2\pi t/T-1 \\
 \cos 2\pi t/T-\sqrt{3} \sin 2\pi t/T-1 & \cos 2\pi t/T+\sqrt{3} \sin 2\pi t/T-1 & -2 \cos 2\pi t/T-1
\end{array}
\right).\nonumber
\ee
\end{widetext}
with $T=2\pi/(\sqrt{3}\omega_0)$. In order to visualize the free dynamics consider $t=T/3$. Then
\be
U_0(T/3)|n,\pm,X_1\rangle
&=&
|n,\pm,X_3\rangle,\\
U_0(T/3)|n,\pm,X_2\rangle
&=&
|n,\pm,X_1\rangle,\\
U_0(T/3)|n,\pm,X_3\rangle
&=&
|n,\pm,X_2\rangle.
\ee
The rotation is counterclockwise, as in Fig.~\ref{Fig0}. In particular, the initial superposition $|\Psi(0)\rangle=\alpha|0,+,X_1\rangle+\beta e^{i\phi}|0,+,X_2\rangle$, prepared in the region where $E(X)=0$, would propagate through the interaction zone as follows,
\be
U_0(T/3)|\Psi(0)\rangle
&=&
\alpha|0,+,X_3\rangle+\beta e^{i\phi}|0,+,X_1\rangle,\\
U_0(2T/3)|\Psi(0)\rangle
&=&
\alpha|0,+,X_2\rangle+\beta e^{i\phi}|0,+,X_3\rangle,\\
U_0(T)|\Psi(0)\rangle
&=&
\alpha|0,+,X_1\rangle+\beta e^{i\phi}|0,+,X_2\rangle.
\ee
$T$ is here an analogue of the time of flight employed in the experiment. 

This would be the case of a free evolution. However, when the interaction at $X_3$ occurs, the dynamics becomes much more interesting. 
The full evolution $U(t)=e^{-i\Omega t}$ is generated by
\be
\Omega
&=&
i \omega_0 \mathbb{I}_2\otimes \mathbb{I}_2\otimes \Big(|X_1\rangle\langle X_2|+|X_2\rangle\langle X_3|+|X_3\rangle\langle X_1|\Big)
\nonumber\\
&\pp=&
-
i \omega_0 \mathbb{I}_2\otimes \mathbb{I}_2\otimes \Big(|X_2\rangle\langle X_1|+|X_3\rangle\langle X_2|+|X_1\rangle\langle X_3|\Big)
\nonumber\\
&\pp=&
+
\omega_1\sigma_x\otimes \sigma_x\otimes |X_3\rangle\langle X_3|.
\ee
In the basis $|n,\pm,j\rangle$ the Hamiltonian is represented by the matrix 
\be
\Omega
&=&
\left(
\begin{array}{cccccccccccc}
 0 & i \omega _0 & -i \omega _0 & 0 & 0 & 0 & 0 & 0 & 0 & 0 & 0 & 0 \\
 -i \omega _0 & 0 & i \omega _0 & 0 & 0 & 0 & 0 & 0 & 0 & 0 & 0 & 0 \\
 i \omega _0 & -i \omega _0 & 0 & 0 & 0 & 0 & 0 & 0 & 0 & 0 & 0 & \omega _1 \\
 0 & 0 & 0 & 0 & i \omega _0 & -i \omega _0 & 0 & 0 & 0 & 0 & 0 & 0 \\
 0 & 0 & 0 & -i \omega _0 & 0 & i \omega _0 & 0 & 0 & 0 & 0 & 0 & 0 \\
 0 & 0 & 0 & i \omega _0 & -i \omega _0 & 0 & 0 & 0 & \omega _1 & 0 & 0 & 0 \\
 0 & 0 & 0 & 0 & 0 & 0 & 0 & i \omega _0 & -i \omega _0 & 0 & 0 & 0 \\
 0 & 0 & 0 & 0 & 0 & 0 & -i \omega _0 & 0 & i \omega _0 & 0 & 0 & 0 \\
 0 & 0 & 0 & 0 & 0 & \omega _1 & i \omega _0 & -i \omega _0 & 0 & 0 & 0 & 0 \\
 0 & 0 & 0 & 0 & 0 & 0 & 0 & 0 & 0 & 0 & i \omega _0 & -i \omega _0 \\
 0 & 0 & 0 & 0 & 0 & 0 & 0 & 0 & 0 & -i \omega _0 & 0 & i \omega _0 \\
 0 & 0 & \omega _1 & 0 & 0 & 0 & 0 & 0 & 0 & i \omega _0 & -i \omega _0 & 0
\end{array}
\right).
\ee
A general state $|\Psi\rangle$ and the initial condition $|\Psi(0)\rangle$ are in this basis given by
\be
|\Psi\rangle
=
\left(
\begin{array}{c}
\Psi_{0-1}\\
\Psi_{0-2}\\
\Psi_{0-3}\\
\Psi_{0+1}\\
\Psi_{0+2}\\
\Psi_{0+3}\\
\Psi_{1-1}\\
\Psi_{1-2}\\
\Psi_{1-3}\\
\Psi_{1+1}\\
\Psi_{1+2}\\
\Psi_{1+3}\\
\end{array}
\right),
\quad
|\Psi(0)\rangle
=
\left(
\begin{array}{c}
0\\
0\\
0\\
\alpha\\
\beta e^{i\phi}\\
0\\
0\\
0\\
0\\
0\\
0\\
0\\
\end{array}
\right)
\ee
($\alpha,\beta,\phi$ are assumed to be real). With this initial condition the problem is effectively six dimensional. It can be further simplified by bringing $\Omega$  to a block-diagonal form by means of  $V$ which diagonalizes $\sigma_x=V^\dag \sigma_z V$.
So, let $W=V\otimes V\otimes \mathbb{I}_3$. Then
\be
U(t)
=
W^\dag\left(
\begin{array}{cccc}
e^{-i\Omega_+t} & 0 & 0 & 0\\
0 & e^{-i\Omega_-t} & 0 & 0\\
0 & 0 & e^{-i\Omega_-t} & 0\\
0 & 0 & 0 & e^{-i\Omega_+t}
\end{array}
\right)
W,
\ee
where
\be
\Omega_\pm
&=&
\left(
\begin{array}{cccc}
0 &  i\omega_0 & -i\omega_0\\
-i\omega_0 & 0 & i\omega_0\\
i\omega_0 & - i\omega_0 & \pm \omega_1
\end{array}
\right).
\ee
\begin{figure}
\includegraphics[width=8 cm]{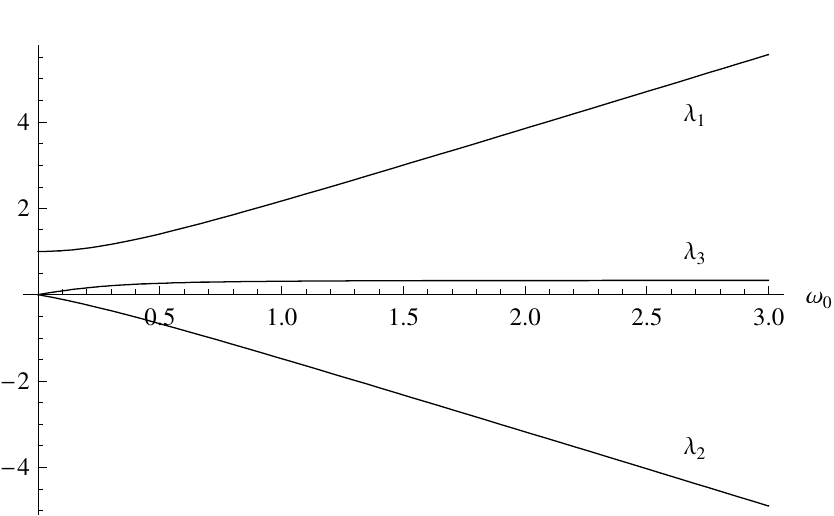}
\caption{Eigenvalues $\lambda_1$, $\lambda_2$, $\lambda_3$, of $\Omega_+$ as functions of $\omega_0$ for $\omega_1=1$.}
\label{Fig lambda}
\end{figure}
The above model is exactly solvable for any $\omega_0$ and $\omega_1$. Let us choose the units of frequency so that $\omega_1=1$. Normalized eigenvectors of 
$\Omega_+=\Omega_+(\omega_0)$ then read
\be
|\lambda\rangle
&=&
\frac{1}{\sqrt{\lambda ^4+3 \omega _0^4}}
\left(
\begin{array}{c}
 -\omega _0^2-i \lambda  \omega _0\\
i \lambda  \omega _0-\omega _0^2\\
\lambda ^2-\omega _0^2
\end{array}
\right),
\ee
where $\lambda$ is one of the three eigenvalues (Fig.~\ref{Fig lambda}),
\begin{widetext}
\be
\lambda_1 &=&\frac{1}{3}+\frac{2}{3} \sqrt{9 \omega _0^2+1} \cos \left(\frac{1}{3} \arg \left(\sqrt{4-4 \left(9 \omega _0^2+1\right){}^3}+2\right)\right),\\
\lambda_2 &=&
\frac{1}{3}-\frac{1}{3} \sqrt{9 \omega _0^2+1} \cos \left(\frac{1}{3} \arg \left(\sqrt{4-4 \left(9 \omega _0^2+1\right){}^3}+2\right)\right)-\frac{\sin \left(\frac{1}{3} \arg
   \left(\sqrt{4-4 \left(9 \omega _0^2+1\right){}^3}+2\right)\right) \sqrt{9 \omega _0^2+1}}{\sqrt{3}},\nonumber\\
\lambda_3 &=&
\frac{1}{3}-\frac{1}{3} \sqrt{9 \omega _0^2+1} \cos \left(\frac{1}{3} \arg \left(\sqrt{4-4 \left(9 \omega _0^2+1\right){}^3}+2\right)\right)+\frac{\sin \left(\frac{1}{3} \arg
   \left(\sqrt{4-4 \left(9 \omega _0^2+1\right){}^3}+2\right)\right) \sqrt{9 \omega _0^2+1}}{\sqrt{3}}.\nonumber
\ee
\end{widetext}
The results for $\Omega_-$ are obtained from  
\be
\Omega_-(\omega_0)=-\Omega_+(-\omega_0).
\ee
Unfortunately, I have not managed to find a value of $\omega_0$ that would make the size of the explicit form of $|\Psi(t)\rangle$ reasonably compact (still, see the next two Sections).
So, let us illustrate the prediction for $\omega_0=\omega_1=1$.  The solution $|\Psi(t)\rangle=e^{-i\Omega t}|\Psi(0)\rangle$ has six vanishing components, 
$\Psi(t)_{0-1}=\Psi(t)_{0-2}=\Psi(t)_{0-3}=\Psi(t)_{1+1}=\Psi(t)_{1+2}=\Psi(t)_{1+3}=0$. The probability of emitting a photon has a Youngian form,
\be
p(t) &=&|\Psi(t)_{1-1}|^2+|\Psi(t)_{1-2}|^2+|\Psi(t)_{1-3}|^2\label{pt}\\
&=& A(t)\alpha^2 +B(t)\beta^2 +C(t)\alpha\beta\cos\phi \label{ptabc}
\ee
and is plotted in Fig.~\ref{Fig1}, for $\alpha=\beta=1/\sqrt{2}$, as a function of both  $\phi$ and $t$. 
The functions $A(t)$ $B(t)$, $C(t)$ from (\ref{ptabc}) are plotted in Fig.~\ref{Fig2'}. 

Let us note that our model has been simplified to its extremes, so that it is clear that the presence of $\cos\phi$ is not a consequence of a Young interference of fields emitted from two different points in space --- the emission here is restricted to a single point $X_3$. It makes no sense to analyze the experiment in terms of which-way measurements either --- the center-of-mass wave packet arrives through a single path. The dependence on $\phi$ is present as well if one restricts the dynamics to one single cycle of evolution, for $0\leq t\leq T=2\pi/\sqrt{3}$ in Fig.~\ref{Fig1}, so that each of the peaks occurs in the interaction zone only once.

\begin{figure}
\includegraphics[width=8 cm]{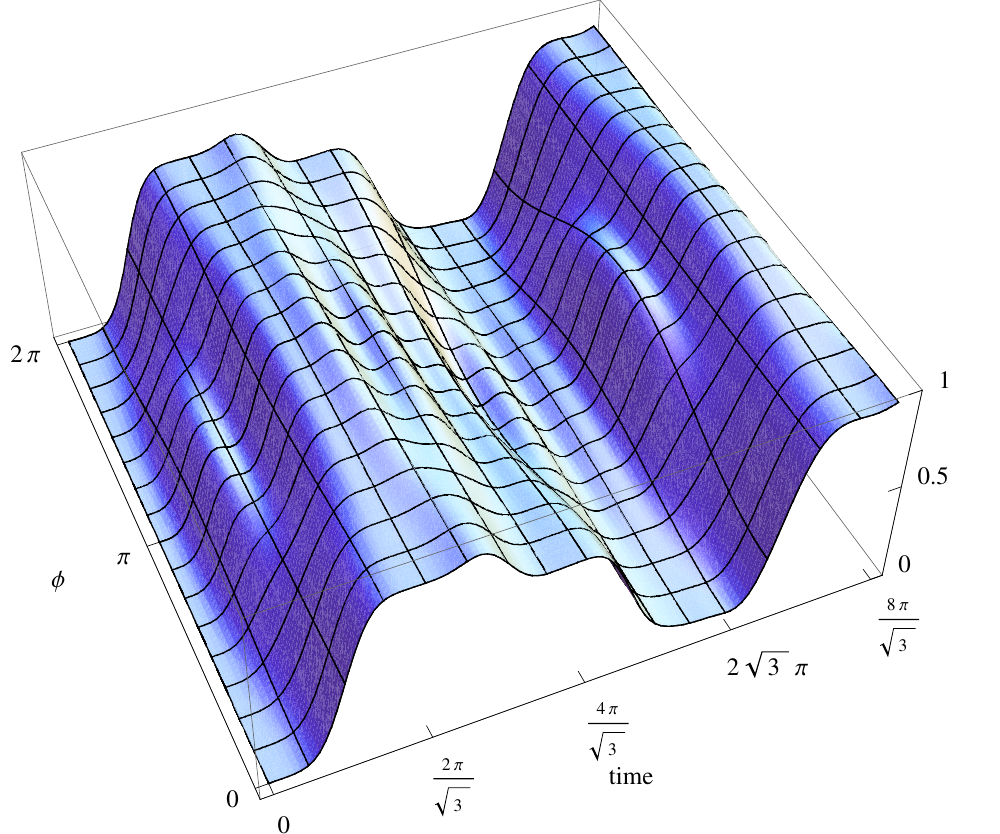}
\caption{Probability (\ref{pt}) as a function of time $t$ and the center-of-mass phase $\phi$, for $\alpha=\beta=1/\sqrt{2}$, $\omega_0=\omega_1=1$. The dependence on $\phi$ is nontrivial and has the same periodicity as the atomic center-of-mass phase, similarly to a Youngian interference pattern. However, both atomic peaks arrive through the same path, and the 
self-interfering photons are emitted from the same point in space. }
\label{Fig1}
\end{figure}
\begin{figure}
\includegraphics[width=8 cm]{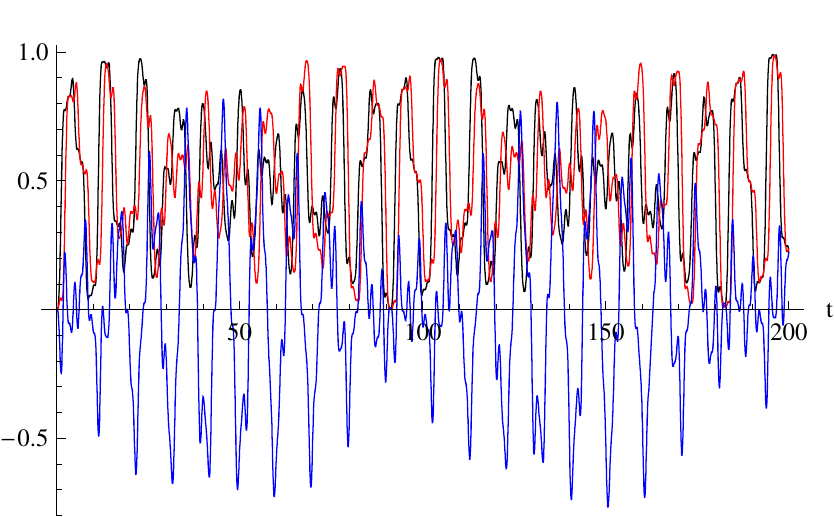}
\caption{The three time-dependent functions occurring in (\ref{ptabc}) for $0\leq t\leq 200$: $A(t)$ (black), $B(t)$ (red), $C(t)$ (blue). The blue curve is a measure of self-interference in time.}
\label{Fig2'}
\end{figure}

\section{Finite-state analogue of a combined longitudinal/transverse Stern-Gerlach interferometer}

The next example is a toy model of two two-peaked wave packets interfering at the interaction zone. This is an analogue of an experiment where one first splits the atomic wave packet by a beam splitter, and then at each of the resulting paths one creates a longitudinal superposition. So effectively, we have here an analogue of a four-peak single-atom wave packet. In the model,  we first observe an interference of $|1\rangle$ arriving at $X_3$ clockwise with 
$|2\rangle$ arriving anticlockwise; then $|2\rangle$ arriving clockwise interferes with $|1\rangle$ arriving anticlockwise. 
Self-interference occurs here in both space and time. 
This can be achieved by taking the free Hamiltonian in the form
\be
\Omega_0
&=&
\omega_0 \mathbb{I}_2\otimes \mathbb{I}_2\otimes \Big(|X_1\rangle\langle X_2|+|X_2\rangle\langle X_3|+|X_3\rangle\langle X_1|\Big)
\nonumber\\
&\pp=&+\textrm{H.c.}
\ee
In spite of a cosmetic change in the free Hamiltonian, the model is mathematically more tractable than the previous one, so a compact form of a solution can be explicitly written. 
The free field evolution operator in position space is no longer a rotation, but a superposition of two opposite rotations:
\begin{widetext}
\be
U_0(t) &=& e^{-i\Omega_0t}
=
-\frac{1}{3}\mathbb{I}_2\otimes \mathbb{I}_2\otimes 
\left(
\begin{array}{lll}
 -2 e^{i t \omega _0}-e^{-2 i t \omega _0} & e^{-2 i t \omega _0} \left(-1+e^{3 i t \omega _0}\right) & e^{-2 i t \omega _0} \left(-1+e^{3 i t \omega _0}\right) \\
 e^{-2 i t \omega _0} \left(-1+e^{3 i t \omega _0}\right) & -2 e^{i t \omega _0}-e^{-2 i t \omega _0} & e^{-2 i t \omega _0} \left(-1+e^{3 i t \omega _0}\right) \\
 e^{-2 i t \omega _0} \left(-1+e^{3 i t \omega _0}\right) & e^{-2 i t \omega _0} \left(-1+e^{3 i t \omega _0}\right) & -2 e^{i t \omega _0}-e^{-2 i t \omega _0}
\end{array}
\right).
\ee
A single cycle of the free dynamics is $T=2\pi/(3\omega_0)$. Taking the same interaction and initial condition as in the previous section, and choosing the same parameters in the Hamiltonian, $\omega_0=\omega_1=1$, 
we find the solution of $i|\dot\Psi(t)\rangle=\Omega|\Psi(t)\rangle$,
\be
|\Psi(t)\rangle
&=&
 \frac{1}{2} e^{i t} \left(\alpha -e^{i \phi } \beta \right)
\left(
\begin{array}{c}
 0 \\
 0 \\
 0 \\
1 \\
 -1\\
 0 \\
 0 \\
 0 \\
 0 \\
 0 \\
 0 \\
 0
\end{array}
\right)
+
\frac{1}{12}(\alpha +e^{i \phi } \beta)
\left(
\begin{array}{c}
 0 \\
 0 \\
 0 \\
 3 e^{-i t} \cos \sqrt{2} t+3 \cos \sqrt{3} t-i \sqrt{3} \sin \sqrt{3} t \\
 3 e^{-i t} \cos \sqrt{2} t+3 \cos \sqrt{3} t-i \sqrt{3} \sin \sqrt{3} t \\
 -i \left(3 \sqrt{2} e^{-i t} \sin \sqrt{2} t+2 \sqrt{3} \sin \sqrt{3} t\right) \\
 3 e^{-i t} \cos \sqrt{2} t-3 \cos \sqrt{3} t+i \sqrt{3} \sin \sqrt{3} t \\
 3 e^{-i t} \cos \sqrt{2} t-3 \cos \sqrt{3} t+i \sqrt{3} \sin \sqrt{3} t \\
 -i \left(3 \sqrt{2} e^{-i t} \sin \sqrt{2} t-2 \sqrt{3} \sin \sqrt{3} t\right) \\
 0 \\
 0 \\
 0
\end{array}
\right).
\ee
\end{widetext}
The explicit form of the Hamiltonian is
\be
\Omega
=\left(
\begin{array}{llllllllllll}
 0 & 1 & 1 & 0 & 0 & 0 & 0 & 0 & 0 & 0 & 0 & 0 \\
 1 & 0 & 1 & 0 & 0 & 0 & 0 & 0 & 0 & 0 & 0 & 0 \\
 1 & 1 & 0 & 0 & 0 & 0 & 0 & 0 & 0 & 0 & 0 & 1 \\
 0 & 0 & 0 & 0 & 1 & 1 & 0 & 0 & 0 & 0 & 0 & 0 \\
 0 & 0 & 0 & 1 & 0 & 1 & 0 & 0 & 0 & 0 & 0 & 0 \\
 0 & 0 & 0 & 1 & 1 & 0 & 0 & 0 & 1 & 0 & 0 & 0 \\
 0 & 0 & 0 & 0 & 0 & 0 & 0 & 1 & 1 & 0 & 0 & 0 \\
 0 & 0 & 0 & 0 & 0 & 0 & 1 & 0 & 1 & 0 & 0 & 0 \\
 0 & 0 & 0 & 0 & 0 & 1 & 1 & 1 & 0 & 0 & 0 & 0 \\
 0 & 0 & 0 & 0 & 0 & 0 & 0 & 0 & 0 & 0 & 1 & 1 \\
 0 & 0 & 0 & 0 & 0 & 0 & 0 & 0 & 0 & 1 & 0 & 1 \\
 0 & 0 & 1 & 0 & 0 & 0 & 0 & 0 & 0 & 1 & 1 & 0
\end{array}
\right).
\ee
Probability (\ref{pt}) now has the form
\be
p(t)=|\alpha+\beta e^{i\phi}|^2 f(t),\label{pt'}
\ee
with $f(t)$ independent of $\phi$. The emission is completely blocked for $\alpha=\beta=1/\sqrt{2}$, $\phi=\pi$. 

\section{The case of a two-peaked source of radiation}

It the previous two examples the interaction was controlled by $X_3$, so the source of radiation was located at a single point in space. Let us now consider the case of four positions: $X_1$ and $X_2$ playing the same role as before, and $X_3$ and $X_4$ controlling the interaction. The dimension of the Hilbert space is $2\times 2\times 4=16$. The example will show that the structure of radiation may involve coherent superpositions of contributions arriving from different atomic peaks. In this concrete example we will see that expressions proportional to $\cos \phi$ cancel each other, although they reappear if one postselects the part of data which is correlated with a single peak. The effect is again exactly opposite to what one  might expect on the basis of entanglement-in-space intuitions.

The 16-dimensional Hilbert space is spanned by $|n\rangle\otimes |s\rangle\otimes |X_j\rangle$, $n=0,1$, $s=\pm$, $j=1,2,3,4$. The free Hamiltonian is a 4-dimensional generalization of the example from the previous section,
\begin{widetext}
\be
\Omega_0
=
\omega_0
\mathbb{I}_2
\otimes
\mathbb{I}_2
\otimes
\Big(
|X_4\rangle\langle X_3|+|X_3\rangle\langle X_2|+|X_2\rangle\langle X_1|+|X_1\rangle\langle X_4|
+
|X_3\rangle\langle X_4|+|X_2\rangle\langle X_3|+|X_1\rangle\langle X_2|+|X_4\rangle\langle X_1|
\Big).
\ee
The interaction part is again a two-qubit NOT gate, but controlled by $X_3$ or $X_4$,
\be
\Omega_{1}
&=&
\omega_1
\sigma_x
\otimes
\sigma_x
\otimes
\Big(
|X_3\rangle\langle X_3|
+
|X_4\rangle\langle X_4|
\Big).
\ee
In matrix form the total Hamiltonian reads
\be
\Omega
=
\left(
\begin{array}{llllllllllllllll}
 0 & \omega _0 & 0 & \omega _0 & 0 & 0 & 0 & 0 & 0 & 0 & 0 & 0 & 0 & 0 & 0 & 0 \\
 \omega _0 & 0 & \omega _0 & 0 & 0 & 0 & 0 & 0 & 0 & 0 & 0 & 0 & 0 & 0 & 0 & 0 \\
 0 & \omega _0 & 0 & \omega _0 & 0 & 0 & 0 & 0 & 0 & 0 & 0 & 0 & 0 & 0 & \omega _1 & 0 \\
 \omega _0 & 0 & \omega _0 & 0 & 0 & 0 & 0 & 0 & 0 & 0 & 0 & 0 & 0 & 0 & 0 & \omega _1 \\
 0 & 0 & 0 & 0 & 0 & \omega _0 & 0 & \omega _0 & 0 & 0 & 0 & 0 & 0 & 0 & 0 & 0 \\
 0 & 0 & 0 & 0 & \omega _0 & 0 & \omega _0 & 0 & 0 & 0 & 0 & 0 & 0 & 0 & 0 & 0 \\
 0 & 0 & 0 & 0 & 0 & \omega _0 & 0 & \omega _0 & 0 & 0 & \omega _1 & 0 & 0 & 0 & 0 & 0 \\
 0 & 0 & 0 & 0 & \omega _0 & 0 & \omega _0 & 0 & 0 & 0 & 0 & \omega _1 & 0 & 0 & 0 & 0 \\
 0 & 0 & 0 & 0 & 0 & 0 & 0 & 0 & 0 & \omega _0 & 0 & \omega _0 & 0 & 0 & 0 & 0 \\
 0 & 0 & 0 & 0 & 0 & 0 & 0 & 0 & \omega _0 & 0 & \omega _0 & 0 & 0 & 0 & 0 & 0 \\
 0 & 0 & 0 & 0 & 0 & 0 & \omega _1 & 0 & 0 & \omega _0 & 0 & \omega _0 & 0 & 0 & 0 & 0 \\
 0 & 0 & 0 & 0 & 0 & 0 & 0 & \omega _1 & \omega _0 & 0 & \omega _0 & 0 & 0 & 0 & 0 & 0 \\
 0 & 0 & 0 & 0 & 0 & 0 & 0 & 0 & 0 & 0 & 0 & 0 & 0 & \omega _0 & 0 & \omega _0 \\
 0 & 0 & 0 & 0 & 0 & 0 & 0 & 0 & 0 & 0 & 0 & 0 & \omega _0 & 0 & \omega _0 & 0 \\
 0 & 0 & \omega _1 & 0 & 0 & 0 & 0 & 0 & 0 & 0 & 0 & 0 & 0 & \omega _0 & 0 & \omega _0 \\
 0 & 0 & 0 & \omega _1 & 0 & 0 & 0 & 0 & 0 & 0 & 0 & 0 & \omega _0 & 0 & \omega _0 & 0
\end{array}
\right).
\ee
\end{widetext}
Diagonalizing $\sigma_x$ we can bring $\Omega$ to a block-diagonal form consisting of four $4\times 4$ blocks. The eigenvalues of $\Omega$ are 
\be
\Omega_{\pm\pm\pm}=\frac{1}{2} \left(\pm 2 \omega _0\pm \omega _1\pm \sqrt{4 \omega _0^2+\omega _1^2}\right),\quad\textrm{etc.}
\ee
where all the eight combinations of pluses and minuses occur, and each eigenvalue is twice degenerate. In order to make the solution as readable as possible we take $\omega_0=2$, $\omega_1=3$.
The initial condition is again a vacuum times a superposition of excited atomic states located at $X_1$ and $X_2$,
\be
|\Psi(0)\rangle
=
\left(
\begin{array}{c}
\Psi_{0-1}\\
\Psi_{0-2}\\
\Psi_{0-3}\\
\Psi_{0-4}\\
\Psi_{0+1}\\
\Psi_{0+2}\\
\Psi_{0+3}\\
\Psi_{0+4}\\
\Psi_{1-1}\\
\Psi_{1-2}\\
\Psi_{1-3}\\
\Psi_{1-4}\\
\Psi_{1+1}\\
\Psi_{1+2}\\
\Psi_{1+3}\\
\Psi_{1+4}
\end{array}
\right)
=
\left(
\begin{array}{c}
0\\
0\\
0\\
0\\
\alpha\\
\beta e^{i\phi}\\
0\\
0\\
0\\
0\\
0\\
0\\
0\\
0\\
0\\
0\\
\end{array}
\right).
\ee
The solution of $i|\dot\Psi(t)\rangle=\Omega|\Psi(t)\rangle$ reads explicitly
\begin{widetext}
\be
\left(
\begin{array}{c}
\Psi_{0-1}\\
\Psi_{0-2}\\
\Psi_{0-3}\\
\Psi_{0-4}\\
\Psi_{0+1}\\
\Psi_{0+2}\\
\Psi_{0+3}\\
\Psi_{0+4}\\
\Psi_{1-1}\\
\Psi_{1-2}\\
\Psi_{1-3}\\
\Psi_{1-4}\\
\Psi_{1+1}\\
\Psi_{1+2}\\
\Psi_{1+3}\\
\Psi_{1+4}
\end{array}
\right)
&=&
\left(
\begin{array}{c}
 0 \\
 0 \\
 0 \\
 0 \\
 \frac{1}{5} (4 \cos t+\cos 4 t) \left(\alpha  \cos 2 t-i e^{i \phi } \beta  \sin 2 t\right) \\
 \frac{1}{5} (4 \cos t+\cos 4 t) \left(e^{i \phi } \beta  \cos 2 t-i \alpha  \sin 2 t\right) \\
 -\frac{2}{5} i \left(e^{i \phi } \beta  \cos 2 t-i \alpha  \sin 2 t\right) (\sin t+\sin 4 t) \\
 -\frac{2}{5} i \left(\alpha  \cos 2 t-i e^{i \phi } \beta  \sin 2 t\right) (\sin t+\sin 4 t) \\
 \frac{8}{5} \left(6 \cos \frac{t}{2}+3 \cos \frac{3 t}{2}+\cos \frac{5 t}{2}\right) \sin ^3\frac{t}{2} \left(i \alpha  \cos 2 t+e^{i \phi } \beta  \sin 2 t\right) \\
 \frac{8}{5} \left(6 \cos \frac{t}{2}+3 \cos \frac{3 t}{2}+\cos \frac{5 t}{2}\right) \sin ^3\frac{t}{2} \left(i e^{i \phi }   \beta  \cos 2 t+\alpha  \sin 2 t\right) \\
\frac{4}{5} (6 \cos t+4 \cos 2 t+2 \cos 3 t+3) \sin ^2\frac{t}{2} \left(i \alpha  \sin 2 t-e^{i \phi } \beta  \cos 2 t\right) \\
\frac{4}{5} (6 \cos t+4 \cos 2 t+2 \cos 3 t+3) \sin ^2\frac{t}{2} \left(i e^{i \phi } \beta  \sin 2 t-\alpha  \cos 2 t\right) \\
 0 \\
 0 \\
 0 \\
 0
\end{array}
\right)
\nonumber\\
&=&
\left(\alpha  \cos 2 t-i e^{i \phi } \beta  \sin 2 t\right)
\left(
\begin{array}{c}
 0 \\
 0 \\
 0 \\
 0 \\
O_{12} \\
0 \\
0 \\
O_{34}\\
I_{12}\\
0\\
0\\
I_{34} \\
 0 \\
 0 \\
 0 \\
 0
\end{array}
\right)
+
\left(e^{i \phi } \beta  \cos 2 t-i \alpha  \sin 2 t\right)
\left(
\begin{array}{c}
 0 \\
 0 \\
 0 \\
 0 \\
0\\
O_{12}  \\
O_{34}\\
0\\
0\\
I_{12}\\
I_{34}\\
0\\
 0 \\
 0 \\
 0 \\
 0
\end{array}
\right)
\ee
where $O_{kl}$ and $I_{kl}$ are time dependent functions defined by the above formula. Now, since
\be
|\alpha  \cos 2 t-i e^{i \phi } \beta  \sin 2 t|^2
&=&
\alpha ^2\cos^2 2t+ \beta ^2\sin^2 2t  +\alpha\beta  \sin 4 t \sin \phi ,\\
|e^{i \phi } \beta  \cos 2 t-i \alpha  \sin 2 t|^2
&=&
\alpha ^2\sin^2 2t+\beta ^2\cos^2 2t  -\alpha\beta  \sin 4 t \sin \phi  ,
\ee
\end{widetext}
the probability of detecting a photon at time $t$,
\be
p(t)
=
\sum_{j=1}^4 |\Psi_{1-j}(t)|^2
=
|I_{12}(t)|^2+|I_{34}(t)|^2,
\ee
is independent of $\phi$. However, probabilities of detecting a photon at time $t$, under the condition that the atom was detected at $X_3$ (or $X_4$ , respectively) are
\be
|\Psi_{1-3}|^2
&=&
|\alpha  \cos 2 t-i e^{i \phi } \beta  \sin 2 t|^2 |I_{34}(t)|^2,\\
|\Psi_{1-4}|^2
&=&
|e^{i \phi } \beta  \cos 2 t-i \alpha  \sin 2 t|^2 |I_{34}(t)|^2.
\ee
Both probabilities depend on $\phi$. The coherence is lost when we do not have the information about the place of emission, but is regained if we know from which peak the photon has arrived. The behavior is completely counterintuitive if one thinks in the categories of the MIT `coherence lost and regained' experiment \cite{Chapman}. 

\section{Can we literally see the atomic phase?}

Young's self-interference is `double-slit but single-time'. Self-interference we have discussed is `single-slit but double-time'.  The models are simplified but exactly solvable, a fact guaranteeing that Youngian terms  are not artifacts of some approximation. However, it is clear that the similarity to the Young effect is superficial and misleading. Interestingly, interference in time is obtained even though time is a parameter. The effect is generic and should be observable in various experimental configurations. For example, the static electric field employed in \cite{1991,1992} could be replaced by a laser beam \cite{Chapman,Mlynek1994,Mlynek1997}. In principle, one should be able to literally see the matter-wave phase by incarnating it into the phase of an emitted radiation.

\section*{Acknowledgments}

It is a great pleasure to  thank Kazimierz Rzążewski for the original inspiration, and David Pritchard for hospitality and patience during my stay in his lab during those memorable years.  I gratefully acknowledge the support I obtained from members of MIT and Paris-Nord atom interferometer groups. Special thanks to Li You for intensive collaboration. My work was partially supported by the Fulbright Commission.

\end{document}